\documentclass[12pt]{article}
\usepackage{amsmath,amssymb}

\usepackage{epsfig,rotating,pifont}

\usepackage{graphicx}
\usepackage{wrapfig}
\usepackage{caption}
\usepackage{subcaption}

\usepackage{hyperref}
\usepackage{cite}

\def\beq{\begin{eqnarray}}
\def\eeq{\end{eqnarray}}

\newcommand{\T}{\textbf{T}}

\newcommand{\bs}{\begin{split}}
\newcommand{\es}{\end{split}}

\newcommand{\F}{{\mathcal{F}}}

\newcommand{\Tr}{\,\mathrm{Tr}\,}            

\newcommand{\be}{\begin{equation}}
\newcommand{\ee}{\end{equation}}
\newcommand{\bea}{\begin{eqnarray}}
\newcommand{\eea}{\end{eqnarray}}
\newcommand{\bg}{\begin{gather}}

\newcommand{\bseq}{\begin{subequations}}
\newcommand{\eseq}{\end{subequations}}

\renewcommand{\ln}{\mathop{\rm ln}\nolimits}

\def\tr{\hbox{Tr}}

\def\be{\begin{eqnarray}}
\def\ee{\end{eqnarray}}
\def\lb{\label}

\begin{document}

\title{\textbf{What surface maximizes 
entanglement entropy?}}

\vspace{2cm}
\author{ \textbf{
Amin Faraji Astaneh$^{1,2,4}$, Gary Gibbons$^{3,4,5}$  and  Sergey N. Solodukhin$^4$ }} 

\date{}
\maketitle
\begin{center}
\hspace{-0mm}
  \emph{$^1$ Department of Physics, Sharif University of Technology,}\\
\emph{P.O. Box 11365-9161, Tehran, Iran}
 \end{center}
 \begin{center}
\hspace{-0mm}
  \emph{$^2$  School of Particles and Accelerators,\\ Institute for Research in Fundamental Sciences (IPM),}\\
\emph{ P.O. Box 19395-5531, Tehran, Iran}
 \end{center}
 \begin{center}
 \hspace{-0mm}
  \emph{$^{3}$ D.A.M.T.P., University of Cambridge, U.K.}
\end{center}
\begin{center}
  \hspace{-0mm}
  \emph{ $^{4}$ Laboratoire de Math\'ematiques et Physique Th\'eorique  CNRS-UMR
7350 }\\
  \emph{F\'ed\'eration Denis Poisson, Universit\'e Fran\c cois-Rabelais Tours,  }\\
  \emph{Parc de Grandmont, 37200 Tours, France} 
  \begin{center}
  \emph{$^5$ LE STUDIUM, Loire Valley Institute for Advanced Studies,}\\
  \emph{ Tours and Orleans, France}
  \end{center}
\end{center}

{\vspace{-11cm}
\begin{flushright}
\end{flushright}
\vspace{11cm}
}



\begin{abstract}
\noindent { For a given  quantum field theory, provided the area of the entangling surface is fixed, what surface maximizes  entanglement entropy?
We analyze the answer to this question in four and higher dimensions.  Surprisingly, in four dimensions the answer  is related to a mathematical problem of 
finding surfaces which minimize  the Willmore (bending) energy and eventually to the Willmore conjecture. We propose a generalization of the Willmore energy in
higher dimensions and analyze its minimizers in a general class of topologies $S^m\times S^n$ and make certain observations and conjectures which may have some mathematical
significance.
}
\end{abstract}

\vskip 2 cm
\noindent
\rule{7.7 cm}{.5 pt}\\
\noindent 
\noindent
\noindent ~~~ {\footnotesize e-mail:  faraji@ipm.ir, gwg1@damtp.cam.ac.uk,  Sergey.Solodukhin@lmpt.univ-tours.fr}

\newpage
    \tableofcontents
\pagebreak

\newpage

\section{Introduction}
\setcounter{equation}0

There are surprisingly many aspects in which entanglement entropy is related to geometry.
This relation is likely to have  very deep reasons which are not yet fully understood.
In this paper we reveal a new aspect of such a relation. It is related to the  problem of finding a surface which maximizes the entanglement entropy
provided the area of the entangling surface is fixed. This problem has many mathematical analogs.  
The particular problem which sparked  our interest is the problem related to sandpiles.

Indeed, suppose $\gamma$ is a curve which bounds the sandpile. If one thinks of the vertical direction
as time, the sand pile may be thought of as the
domain of dependence of the base of the
the sand pile. One may ask, for a given perimeter $\ell(\gamma)$, what sandpile has the largest volume.
The answer, perhaps unsurprisingly, is when $\gamma$ is a circle \cite{sandpile}. In the holographic approach to the entanglement entropy
the geometrical picture is pretty much similar to that of the sandpile so that one would expect that something similar is going on.

In this paper we make this intuitive picture more precise.  As in the case of the sandpiles the entanglement entropy is indeed
maximized by the round sphere (in dimensions $d\geq 4$).  So that the round sphere is what we shall call the global entropy maximizer.
However, in each topological class there may exist its own entropy maximizer and indeed, as we show in this paper, this is the case.
In four dimensions these maximizers are the so-called Lawson surfaces, higher genus compact surfaces which can be minimally embedded in sphere $S^3$.
For genus $g=1$ the surface is the Clifford torus and our problem is related to the so-called Willmore conjecture, the problem of minimization of the
Willmore energy. 

In higher dimensions the situation is more complicated since the topological classification of compact $(d-2)$-dimensional surfaces is more involved.
We, nevertheless, analyze this problem for some particular class of surfaces which have the product structure $S^m\times S^n$, $n+m=d-2$, and find the respective entropy 
maximizers. Moreover, we present  arguments why the round sphere has to be the global entropy maximizer. This therefore answers the question made in the title of the paper.

\section{Preliminaries}
\setcounter{equation}0
\subsection{Entanglement entropy}
Consider a  $d$-dimensional spacetime $\cal M$ and a co-dimension  two surface $\Sigma$.
For a given compact closed surface $\Sigma$ entanglement entropy is defined by tracing over
degrees of freedom resigning inside the surface. Provided one starts with a pure quantum (typically vacuum) 
state, after the tracing one ends up with a non-trivial density matrix $\rho$. Entanglement entropy (for a review see \cite{EE}) is then defined as
\be
S(\Sigma)=-\Tr\rho(\Sigma) \ln\rho(\Sigma)\, .
\lb{1}
\ee
Remarkably, tracing over degrees of freedom outside the surface gives the same value. This property of the entropy 
indicates that entanglement entropy is not an extensive quantity which is characterized by geometry of the surface
and the spacetime near the surface. One  quantity which encodes geometry is the area of the surface, $A(\Sigma)$.
Indeed, being computed in a quantum field theory, to leading order entanglement entropy is found to be proportional to the area.
Since entropy is a dimensionless quantity its dependence on the area should be compensated by  another dimensionfull variable.
This variable naturally appears in a  quantum field theory and is known as a UV cut-off $\epsilon$.  
Thus, entanglement entropy is a function of the UV cut-off $\epsilon$ as well as of the geometric characteristics of the surface,
$S=S(\Sigma, \epsilon)$. Now, we are ready to formulate our problem.

\subsection{Formulation of the problem}
\lb{formulation}
{\it (A) Suppose that the quantum field theory is specified and fixed as well as the background spacetime $\cal M$.
In particular, the UV cut-off $\epsilon$ is fixed so that the entropy can be considered to be function only of geometry
of the surface $\Sigma$, $S(\Sigma)$. Consider a class of surfaces of the same area, $A=A(\Sigma)$. This class may be also specified
by imposing certain restrictions on topology of $\Sigma$. For what surface  $\Sigma_0$ does the entanglement entropy $S(\Sigma)$ 
take the maximal value?}

If, for a given topology, such a surface $\Sigma_0$  exists then we obviously have an inequality
\be
S(\Sigma)\leq S(\Sigma_0)\ \ {\it topology \ is \ fixed}
\lb{2}
\ee
We shall call $\Sigma_0$ a ``maximizer'' of the entropy.  A separate interesting problem is to find a global maximizer.

{\it (B) Suppose that all conditions  of (A) hold but the topology of surface is not fixed and can vary.
Is there a surface $\Sigma_m$, called a global ``maximizer'', such that
\be
S(\Sigma)\leq S(\Sigma_m)\ \ {\it any \ topology}
\lb{3}
\ee
for any surface $\Sigma$ of same area $A$ and arbitrary topology?}

\subsection{Minkowski spacetime: a natural guess for maximizer}
\lb{statement}
In this paper we mostly consider the case when spacetime $\cal M$ is Minkowski.
In $d$-dimensional Minkowski spacetime there is a large group of symmetry $O(d)$. This symmetry may be useful 
in finding a maximizer. Indeed,  a surface-maximizer, $\Sigma_0$ or $\Sigma_m$, is most likely to be a maximally symmetric surface, i.e.
to be invariant under a  group of rotations $O(d-1)$. There is only one such surface, the round sphere $S^{d-2}$.

Therefore, we might  guess that {\it  the round sphere is the maximizer in its own topological class.}  This is one of the conjectures which we shall check in this paper.
 This symmetry consideration, however,  does not tell us whether the round sphere $S^{d-2}$ is a global maximizer and what surfaces maximize the entropy in
 other topological classes which do not contain spheres.
We therefore formulate our proposed answer to question ({\it B}):

\medskip

\noindent {\it The round sphere $S^{d-2}$ is the global maximizer of entanglement entropy in any topology,
\be
S(\Sigma_{d-2})\leq S(S^{d-2})\, .
\lb{sphere}
\ee
}
Below we shall provide evidence for this statement in various dimensions.

\subsection{Holographic entanglement entropy}
One way to attack the problem outlined in section \ref{formulation} is to use the holographic approach to entanglement entropy proposed
in  \cite{Ryu:2006bv}. According to this approach one considers a $(d+1)$-dimensional spacetime which solves Einstein equations with a negative cosmological constant.
This spacetime is asymptotically Anti-de Sitter and we shall use notation $AdS_{d+1}$ even though this space is not globally Anti de Sitter.
The physical $d$-dimensional spacetime $\cal M$ is conformal boundary of 
$AdS_{d+1}$. We remind the reader that the entangling surface $\Sigma$ is co-dimension 2 surface in $\cal M$. Now, in a hypersurface   of constant time
in $AdS_{d+1}$ consider a $(d-1)$-dimensional surface ${\cal H}_\Sigma$ which bounds entangling surface $\Sigma$, $\partial{\cal H}_\Sigma=\Sigma$.
Let us impose condition that ${\cal H}_\Sigma$ to be minimal surface. Its area is ${\cal A}({\cal H}_\Sigma)$. It is divergent and it should be regularized by placing $\cal M$ at some finite small
distance $\epsilon$ from infinity of $AdS_{d+1}$.  In the holographic dictionary $\epsilon$ should be identified with the UV cut-off in a conformal field theory
in physical space $\cal M$. Now, according to prescription of \cite{Ryu:2006bv} the entanglement entropy in the CFT living in $\cal M$ and defined for 
the entangling surface $\Sigma$ is given by 
\be
S_{HE}(\Sigma)=\frac{A({\cal H}_\Sigma)}{4G_{d+1}}\, ,
\lb{3-1}
\ee
where $G_{d+1}$ is $(d+1)$-dimensional Newton's constant. According to the holographic dictionary $G_N$ is related to number of degrees of freedom in the CFT.
For instance $1/G_3=2/3N$ ($d=2$), $1/G_5=2/\pi N^2$ ($d=4$), $1/G_7=32/\pi^2 N^3$ ($d=6$) so that for generic $d$, $1/G_{d+1}\sim N^{d/2}$.

Asymptotically, near the conformal boundary of $AdS_{d+1}$ the equation for a  minimal surface ${\cal H}_\Sigma$ 
can be found by using the Fefferman-Graham coordinates. In four dimensions the analysis was done by Graham and Witten \cite{Graham:1999pm}.
Generalizing this analysis for arbitrary dimensions $d$ (see also \cite{Myers:2013lva} for a relevant analysis) we find 
the following asymptotic  for the volume element of the  minimal hypersurface, $\mathcal{H}_\Sigma$
\begin{equation}
dv_{\mathcal{H}_\Sigma}=r^{-d+1}\left[1-\frac{1}{2}\left(\frac{d-3}{(d-2)^2}(\Tr K)^2+\Tr P\right)r^2+\cdots\right]dv_\Sigma\  dr\ ,
\lb{3-2}
\end{equation}
where $r$ is the radial coordinate orthogonal to $\cal M$ and  following the notation of Graham-Witten we have
\begin{equation}
P_{\alpha\beta}=\frac{1}{d-2}\left(R_{\alpha\beta}-\frac{R}{2(d-1)}g_{\alpha\beta}\right)\ .
\lb{3-3}
\end{equation}
$R_{\alpha\beta}$ here is the curvature in physical spacetime $\cal M$ and $K^a_{\alpha\beta}$, $a=1,\, 2$ is extrinsic curvature of surface $\Sigma$.
The traces in  (\ref{3-2}) are defined in terms of the induced metric on $\Sigma$, $\gamma_{\alpha\beta}=g_{\alpha\beta}-\Sigma_{a=1,2}(n^a_{\alpha}n^a_\beta)$. We then 
arrive at 
\begin{equation}
\Tr P=\frac{1}{d-2}\left(-R_{aa}+\frac{d}{2(d-1)}R\right)\ 
\end{equation}
$R_{aa}=R_{\alpha\beta} n^\alpha_a n^\beta_a$.
Now putting things together and performing the integral over $r$ (which goes from $r=\epsilon$ on the lower limit) one finds the asymptotic form for the holographic entanglement entropy 
\begin{equation}\label{3-4}
\begin{split}
S_{HE}(\Sigma)&=\frac{A({\mathcal{H}_\Sigma})}{4G_{d+1}}=\frac{1}{4G_{d+1}}\frac{A(\Sigma)}{(d-2)\epsilon^{d-2}}+\\
&+\frac{1}{4G_N}\frac{1}{2(d-2)(d-4)\epsilon^{d-4}}\int_\Sigma dv_\Sigma \left[R_{aa}-\frac{d}{2(d-1)}R-\frac{d-3}{d-2}(\Tr K)^2\right]\ .
\end{split}
\end{equation}
In dimension $d=4$  the power law divergence $1/\epsilon^{d-4}$ becomes logarithmic and we arrive at
\begin{equation}
S_{HE}(\Sigma)=\frac{A(\Sigma)}{4\pi\epsilon^2}N^2+\frac{N^2}{24\pi}\int_\Sigma dv_\Sigma\, (3R_{aa}-2R-\frac{3}{2}(\Tr K)^2)\log\frac{1}{\epsilon}  
\lb{3-5}
\end{equation}
in agreement with earlier derivation in \cite{Solodukhin:2008dh} (see also \cite{Solodukhin:2006xv} for the case of vanishing extrinsic curvature). Expression (\ref{3-4}) agrees with the holographic calculation for a $(d-2)$-sphere
given in \cite{Ryu:2006bv}.

\section{ Maximum of entropy in dimension $d=4$}
\setcounter{equation}0
\subsection{Holographic analysis}
Let us consider the case when the physical spacetime is flat. Then the holographic formula (\ref{3-5}) simplifies
\be
S_{HE}(\Sigma)=\frac{A(\Sigma)}{4\pi\epsilon^2}N^2-\frac{N^2}{16\pi}\int_\Sigma dv_\Sigma\, (\Tr K)^2\, \log\frac{1}{\epsilon}\, , 
\lb{4-1}
\ee
so that the only contribution from the  extrinsic curvature is in the logarithmic term.

In our problem formulated in section \ref{formulation} one considers  a class of surfaces of same area $A=A(\Sigma)$. Thus, in this class
the first term in (\ref{3-5}) is the same for all surfaces.  Therefore, in order to find a maximum of the entropy (\ref{3-5}) one has to look at the {\it minimum}
of the second term which is proportional to the integral of square of extrinsic curvature. This term is well known in the mathematical literature as the Willmore (bending) energy
\be
W(\Sigma)=\frac{1}{4} \int_\Sigma (\Tr K)^2\, .
\lb{W}
\ee
Analyzing its minimum we first do some rewriting
\be
\frac{1}{2}(\Tr K)^2&=&R_\Sigma+K_\Sigma\, , \nonumber \\
R_\Sigma=(\Tr K)^2-\Tr K^2\, &,& \, K_\Sigma=\Tr K^2-\frac{1}{2}(\Tr K)^2\, ,
\lb{4-3}
\ee
where $R_\Sigma$ is intrinsic curvature of $\Sigma$. The important observation now is that invariant $K_\Sigma$ is a complete square
\be
K_\Sigma=(K_{ij}-\frac{1}{2}\gamma_{ij}\tr K)^2\, ,
\lb{4-4} 
\ee
where $\gamma_{ij}$ is the induced metric on $\Sigma$. Since the first term in (\ref{4-3}) is topological the minimum of (\ref{4-3}) is achieved if the second term, integral of a complete square (\ref{4-4}), vanishes.
This is only possible if $K_\Sigma=0$, i.e.
\be
K_{ij}=\frac{1}{2}\gamma_{ij}\Tr K\, .
\lb{4-5}
\ee
Using the Gauss-Codazzi equations
\be
&&\nabla^j K_{ij}=\nabla_i \tr K\, ,\nonumber \\
&&R_{\Sigma}=(\tr K)^2-\tr K^2\, 
\lb{4-6}
\ee
we find that eq.(\ref{4-5}) implies that $R_\Sigma$ is constant and positive so that $\Sigma$ is a round sphere $S^2$.

Thus, we have proved that the round sphere is the minimizer of the Willmore energy. This is of course a well known fact in the  mathematical literature.
For us it implies that, in its topological class, the round sphere  is the maximizer of the holographic entanglement entropy
\be
S_{HE}(\Sigma)\leq S_{HE}(S^2)\, .
\lb{4-7}
\ee

\subsection{Generic 4d CFT in Minkowski spacetime}
The  above holographic consideration can be generalized to cover  the entanglement entropy of a generic 4d conformal field theory.
Indeed, the UV divergent terms in the entropy  read \cite{Solodukhin:2008dh}
\be
S_{CFT}(\Sigma)=\frac{{N}(a,b)\, Area(\Sigma)}{4\pi\epsilon^2}-\frac{1}{2\pi}\left(a\int_\Sigma R_\Sigma+b\int_\Sigma K_\Sigma \right)\ln\frac{1}{\epsilon}\, ,
\lb{4-8}
\ee
where $N(a,b)$ is the number of on-shell degrees of freedom in the CFT, $a$ and $b$ are central charges related to conformal anomalies. For free fields we have that
the a-charge is non-negative, $a\geq 0$,  and $b$-charge is positive,  $b>0$, for all fields except  $s=3/2$. 
The CFT holographically dual to supergravity on $AdS_5$ is characterized by condition $a=b$.
Provided the spectrum of the CFT does not contain exotic particles, the  same arguments as above apply to this more general case.

By the same arguments  as before entropy (\ref{4-8}) has maximum for a surface $\Sigma_0$ for which $K_\Sigma=0$ so that $\Sigma$ is a round sphere, $\Sigma_0=S^2$.
Hence, we still have the bound
\be
S_{CFT}(\Sigma)\leq S_{CFT}(S^2)\, .
\lb{4-9}
\ee

\subsection{A mass deformation of CFT}
These  consideration can be even generalized to non-conformal theories. Consider  a deformation of the CFT by adding some mass. Then the entanglement entropy 
takes the form \cite{Fursaev:2013fta}
\be
S_{non-CFT}(\Sigma)=\frac{N(a,b)\, A(\Sigma)}{4\pi\epsilon^2}\, -\, \frac{1}{2\pi}\int_\Sigma\left(a R_\Sigma+b K_\Sigma +\sum_s \frac{m_s^2}{12}D_s\right)\ln\frac{1}{\epsilon}\, ,
\lb{4-10}
\ee
where $D_s$ is the dimension of representation of spin $s$. The mass term in the entropy (\ref{4-10}) again reduces to an area term and it is thus irrelevant for the 
maximization of the entropy.  The maximum  is  then achieved for a surface for which $K_\Sigma=0$ so that this surface is again the round sphere,
\be
S_{non-CFT}(\Sigma)\leq S_{non-CFT}(S^2)\, .
\lb{4-11}
\ee
This inequality in Minkowski spacetime is thus quite robust and is valid for a very large class of theories.

\subsection{Curved space-time}

For a 4d  CFT in curved space-time the entanglement entropy is modified by the Weyl tensor projected on the sub-space transverse to $\Sigma$ (see \cite{Solodukhin:2008dh}, \cite{Solodukhin:2011zr})
\be
S(\Sigma)=\frac{N(a,b)\, Area(\Sigma)}{4\pi\epsilon^2}\
-\frac{1}{2\pi}\int_\Sigma\left(a R_\Sigma+b(-W_{abab}+ K_\Sigma) \right)\ln\frac{1}{\epsilon}\, ,
\lb{4-12}
\ee
where $K_\Sigma$ is still defined as in (\ref{4-3}). The round sphere is still a maximizer of the entropy if Weyl tensor of spacetime $\cal M$ is vanishing.
It would be interesting to generalize this to spacetime with a non-trivial Weyl tensor.

\subsection{Higher genus: the Willmore conjecture and the Lawson surfaces}
In four dimensions, $d=4$, as we have shown in this paper, the problem of maximizing entanglement entropy is related to a mathematical problem of minimization
of the Willmore bending energy (\ref{W}). The topological type of 2-dimensional closed surface $\Sigma$ is completely
characterized by its genus $g$. If $g=0$ (sphere) the minimizer of the Willmore energy  is round sphere $S^2$ for which $W(S^2)=4\pi$.
This is consistent with our consideration above. But what about higher genus? The answer to this question is given by the so-called Willmore conjecture.
For genus $g=1$ (torus) Willmore \cite{Willmore} conjectured in 1966 that for surfaces of
higher genus there exists a better bound. In particular, for any torus one has that
\be
W(\T^2)\geq 2\pi^2
\lb{T}
\ee
with equality if and only if $\T^2$ is the Clifford torus. This torus is characterized by the ratio
of two radii equal to $1/\sqrt{2}$. For the Clifford torus the Willmore energy $W(\T^2)=2\pi^2>4\pi$. So that the round sphere is still the energy minimizer
in these two topological classes. The Willmore conjecture  was proven very recently in \cite{proof}. With this conjecture we obtain that for surfaces of  genus $g=1$
entanglement entropy satisfies the bound
\be
S(\Sigma_{g=1})\leq S(\T^2_{Cliff})
\lb{ST}
\ee
with equality if and only if $\Sigma$ is the Clifford torus $\T_{Cliff}^2$.

In order to illustrate the Willmore conjecture  let us consider the 4-dimensional flat metric 
\begin{equation}
ds^2=\left(dt^2+ dr^2+r^2d\theta^2+(R_2+r\cos\theta)^2d\phi^2\right)\, ,
\lb{metric-T}
\end{equation}
where $0\leq\phi, \theta\leq 2\pi$.
In this metric we can define torus as a hypersurface of constant $r=R_1$. In this metric $R_2$ is the radius of the circle  of revolution and $R_1$  is the radius of the circle being revolved.  
Clearly, $R_2\geq R_1$. Effectively, we can discard the time in (\ref{metric-T}) and consider  the 2-torus as immersed in $R^3$. 
The area of the torus is found to be
\begin{equation}
A(\T^2)=\int_0^{2\pi} d\phi\int_0^{2\pi} d\theta \,\, R_1(R_2+R_1\cos\theta)=4\pi^2\,R_1R_2\, .
\end{equation}
The trace of the extrinsic curvature  is found to be
\begin{equation}
K(\T^2)=\frac{R_2+2R_1\cos\theta}{R_1(R_2+R_1\cos\theta)}\, .
\end{equation}
Then one can evaluate the Willmore energy as
\begin{equation}
\begin{split}
W(\T^2)&=\int_0^{2\pi} d\phi\int_0^{2\pi} d\theta \,\,\left[\frac{R_2+2R_1\cos\theta}{2R_1(R_2+R_1\cos\theta)}\right]^2R_1(R_2+R_1\cos\theta)\\
&=\frac{\pi^2R_2^2}{R_1\sqrt{R_2^2-R_1^2}}\, .
\end{split}
\end{equation}
This energy is a function of the ratio $x=R_2/R_1$ and it has a  minimum at $x=\sqrt{2}$ that corresponds to $R_2=\sqrt{2}R_1$. This is exactly the Clifford torus.
 As expected, the Willmore energy for this value of $x$ is
\begin{equation}
W(\T^2)\Big\vert_{R_2=\sqrt{2}R_1}=2\pi^2\, .
\end{equation}

Sometimes, it is convenient, and in fact preferred by mathematicians,  to consider, instead  of flat space $R^3$,   2-surfaces embedded  in sphere $S^3$.  The two spaces are conformally related by 
stereographic projection.
The conformal invariant generalization of the Willmore energy is 
\be
W(\Sigma)=\frac{1}{4} \int_\Sigma \left((\Tr K)^2-R_{abab}+\frac{1}{2}R_{aa}\right)\, .
\lb{W-conf}
\ee
For a 2-surface $\Sigma$ embedded in $S^3$ we have 
\be
W(\Sigma)=\int_\Sigma \left(\frac{1}{4}(\Tr K)^2+1\right)\, .
\lb{Wmin}
\ee
Among all possible embeddings the distinguished one is the minimal embedding. If $\Sigma$ is minimally embedded in $S^3$ the trace of extrinsic curvature of $\Sigma$
vanishes and the Willmore energy 
\be
W(\Sigma)=A(\Sigma)
\lb{W=A}
\ee
is equal to the area of $\Sigma$.  A 2-sphere $S^2$ is minimally embedded in $S^3$ as the equator. On the other hand, the Clifford torus,  which is a square torus defined by equations
\be
x_1^2+x_2^2=\frac{1}{2}=x_3^2+x_4^2\, ,
\lb{CT}
\ee
is the only
torus which can be minimally  embedded in $S^3$. This statement is known as the Lawson conjecture made by H.B. Lawson, Jr. in 1970 \cite{Lawson}.
This conjecture was proven in 2013 by S. Brendle \cite{Brendle}.

In higher genus $g>1$  Lawson has constructed surfaces $\Sigma_{g,L}$ which are minimally embedded in $S^3$ (for a recent review see \cite{Brendle2}). The area of the Lawson surface satisfies inequality
\be
4\pi <A(\Sigma_{g,L})<8\pi\, .
\lb{WL}
\ee
It was conjectured in \cite{KP} that this area is monotonically increasing in $g$ and that in the limit $g\rightarrow \infty$ it is approaching $8\pi$. 
These surfaces, as conjectured by Kusner \cite{Kusner} in 1989, are the minimizers of the Willmore energy in their genus. 
Thus, in our problem of maximization the entanglement entropy, namely the Lawson  surfaces are the  entropy maximizers
in their respective genus,
\be
S(\Sigma_g)\leq S(\Sigma_{g,L})\, .
\lb{SL}
\ee
 On the other hand, the round sphere  is the global maximizer for all   genera,
 \be
 S(\Sigma_g)\leq S(S^2)
 \lb{SS}
 \ee
 for any genus $g$ in agreement with our statement in section \ref{statement}.
 
 This completes our analysis in dimension $d=4$.

\section{ Maximum of entropy in dimension $d>4$}
\setcounter{equation}0
\subsection{Holographic entanglement entropy in higher dimensions}
We continue our analysis for the holographic entanglement entropy in higher dimension $d>4$. 
In flat physical spacetime, as follows from eq.(\ref{3-4}), the leading UV divergent part of the entropy takes the form
\begin{equation}\label{5-1}
\begin{split}
S_{HE}(\Sigma)=\frac{{\cal N} A(\Sigma)}{(d-2)\epsilon^{d-2}}-\frac{{\cal N}(d-3)}{2(d-2)^2(d-4)\epsilon^{d-4}}\int_\Sigma dv_\Sigma (\Tr K)^2\ ,
\end{split}
\end{equation}
where $\cal N$ is an effective number of degrees of freedom in the theory.
The second term in this expression is again proportional to a higher dimensional analog of the Willmore energy. However, in higher dimensions
this quantity should be properly understood as we discuss in the next subsection.

\subsection{Higher dimensions and the normalized Willmore energy}
In higher dimensions the Willmore energy (\ref{W}) is dimensionfull and therefore can be easily made  arbitrary small by just rescaling the size of
of the surface. Therefore, in order to have a sensible minimization problem we need to define a quantity which would be dimensionless. 
A natural generalization is divide the higher dimensional analog of (\ref{W}) by an appropriate power of the surface area $A=A(\Sigma)$.
Thus in dimension $d$ we define the normalized Willmore energy as follows
\be
\widehat{W}(\Sigma_{d-2})=W(\Sigma_{d-2})/A^{\frac{d-4}{d-2}}\, , \,\, \, W(\Sigma_{d-2})=\frac{1}{4}\int_{\Sigma_{d-2}}(\Tr K)^2\, .
\lb{5.1}
\ee
The minimization problem of this quantity will be analyzed later in the paper. The respective minimizers, as will be shown,  are the maximizers of the entanglement entropy
(provided the area of surface is fixed) in the appropriate dimensions. For a recent work on the generalized Willmore energy see \cite{Gover:2014iba}.

\subsection{Round sphere}
We start our analysis with the case when the 
entangling surface $\Sigma$ is a round $(d-2)$-sphere of radius $R$.
The area of the sphere is
\begin{equation}
A(S^{d-2})=\frac{2\pi^\frac{d-1}{2}}{\Gamma\left(\frac{d-1}{2}\right)}R^{d-2}\, ,
\lb{5.2}
\end{equation}
while the  Willmore functional is equal to
\begin{equation}
W(S^{d-2})=\frac{(d-2)^2}{4R^2}A(S^{d-2})=\frac{(d-2)^2}{4}\frac{2\pi^\frac{d-1}{2}}{\Gamma\left(\frac{d-1}{2}\right)}R^{d-4}\, , 
\lb{5.3}
\end{equation}
so that the normalized Willmore energy is
\begin{equation}
\widehat{W}(S^{d-2})=\frac{W(S^{d-2})}{[A(S^{d-2})]^\frac{d-4}{d-2}}=\frac{(d-2)^2}{4}\left(\frac{\pi^\frac{d-1}{2}}{\Gamma\left(\frac{d-1}{2}\right)}\right)^{\frac{2}{d-2}}\, .
\lb{5.4}
\end{equation}
Respectively, as follows from (\ref{5-1}), the entanglement entropy of a round sphere of area $A$ in $d$ dimensions to leading order takes the form
\be
S(S^{d-2})=\frac{{\cal N} A}{(d-2)\epsilon^{d-2}}-\frac{{\cal N}c(d) }{\epsilon^{d-4}}A^{\frac{d-4}{d-2}}\, , \ \ c(d)=\frac{(d-3)\pi^{\frac{d-1}{d-2}}}{2(d-4)(\Gamma(\frac{d-1}{2}))^{\frac{2}{d-2}}}\, .
\lb{5.5}
\ee

\subsection{Ellipsoid}
In order to approach our problem and check whether the round sphere is a maximizer of entropy (and respectively a minimizer of the normalized Willmore energy)
 it is natural to consider a deformation of the sphere which can  still be  treated analytically. An example of a  deformation of this type is 
the ellipsoid. An Ellipsoid $E^{d-2}$ is a surface in space $R^{d-1}$ with Cartesian coordinates $(x_1,x_2,..,x_{d-1})$  described by the equation 
\be
\frac{x_1^2}{a_1^2}+\, ...\, +\frac{x_{d-1}^2}{a_{d-1}^2}=1\, .
\lb{5.5}
\ee
One may choose the angular coordinates $(\theta_1,..,\theta_{d-2})$ as follows
\begin{equation}\label{5.6}
\begin{split}
&x_1=a_1 \cos\theta_1\cos\theta_2\cdots\cos\theta_{d-3}\, ,\\
&x_2=a_2  \sin\theta_1\cos\theta_2\cdots\cos\theta_{d-3}\, ,\\
&x_3=a_3  \sin\theta_2\cdots\cos\theta_{d-3}\, ,\\
&\vdots\\
&x_{d-1}=a_{d-1}\sin\theta_{d-2}\, .
\end{split}
\end{equation}
In what follows for simplicity we shall restrict ourselves to a special case at which
\begin{equation}
(a_1=a_2=\cdots=a_{d-2}= a)\neq (a_{d-1}=b)\, .
\lb{5.7}
\end{equation}
The area element reads
\begin{equation}\lb{5.8}
dv(E^{d-2})=a^{d-3}\cos\theta_2\times\cdots\times(\cos\theta_{d-2})^{d-3}\ \F^\frac{1}{2}(\theta_{d-2})\prod_{i=1}^{d-2} d\theta_i\ ,
\end{equation}
while the trace of the extrinsic curvature 
\begin{equation}
\Tr K(E^{d-2})=\frac{b}{a}\, \frac{a^2+(d-3)\F(\theta_{d-2})}{\F^{\frac{3}{2}}(\theta_{d-2})}\ ,
\lb{5.9}
\end{equation}
where we have defined
\begin{equation}
\F(\theta_{d-2})=(a^2\sin^2\theta_{d-2}+b^2\cos^2\theta_{d-2})\ .
\lb{5.10}
\end{equation}
Now after performing the integrals, the area of the ellipsoid is found to be  
\begin{equation}
A(E^{d-2})=\frac{2\pi^{\frac{d-1}{2}}}{\Gamma\left(\frac{d-1}{2}\right)}a^{d-2}(1-e^2)^\frac{d-2}{2}\ {_2F_1(\frac{d}{2},\frac{d-2}{2},\frac{d-1}{2},e^2)}\ .
\lb{5.11}
\end{equation}
in terms of the hypergeometric function, where we introduced $e=\sqrt{1-\frac{a^2}{b^2}}$.

\begin{figure*}[t]
\centering
\includegraphics[width=1.0\textwidth]{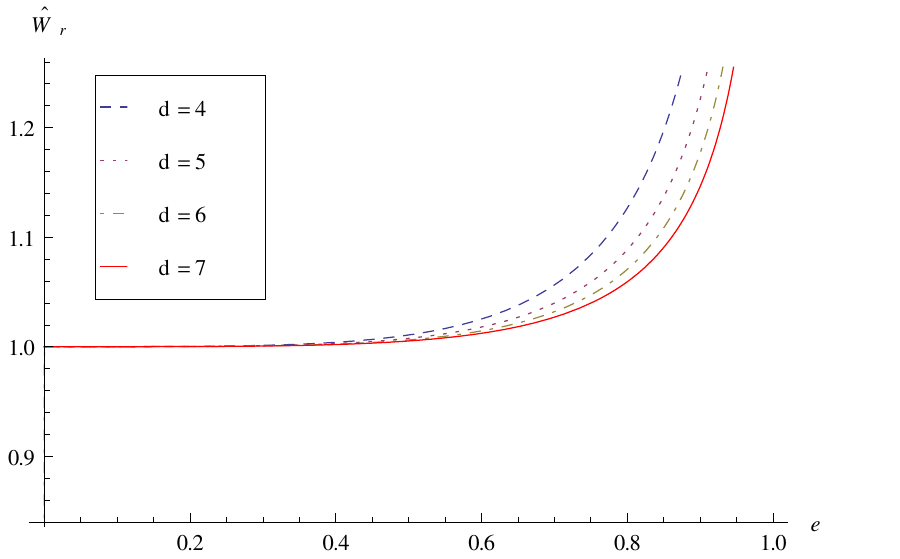}
\caption{Ratio of normalized Willmore energies (ellipsoid to sphere) in  dimension $d$. }\label{fig1}
\end{figure*}

On the other hand,  the Willmore energy of the ellipsoid reads
\begin{equation}
\begin{split}
W(E^{d-2})=
&\frac{1}{4}\frac{2\pi^{\frac{d-1}{2}}}{\Gamma\left(\frac{d-1}{2}\right)}a^{d-4}(1-e^2)^\frac{d-4}{2}
\Bigg[{_2F_1(\frac{d-2}{2},\frac{d-6}{2},\frac{d-1}{2},e^2)}\\
&+(d-3)^2\ {_2F_1(\frac{d-2}{2},\frac{d-2}{2},\frac{d-1}{2},e^2)}\\
&+2(d-3)\ {_2F_1(\frac{d-2}{2},\frac{d-4}{2},\frac{d-1}{2},e^2)}\Bigg]\ .
\end{split}
\lb{5.12}
\end{equation}
We are interested in the ratio of the normalized energies for ellipsoid and sphere,
\be
\widehat{W}_r(e)&=\frac{\widehat{W}(E^{d-2})}{\widehat{W}(S^{d-2})}\, .
\lb{5.13}
\ee
In figure (\ref{fig1}),  we have plotted some curves describing the behavior of this function with respect to the parameter $e$ in various dimensions. 
In all dimensions we find that this is a monotonically growing function. Its minimum is equal to 1 at $e=0$,  corresponding  to the round sphere. At $e=1$ the function (\ref{5.13}) approaches infinity. The case $e=1$ corresponds to $a=0$ so that the ellipsoid shrinks to an interval. Thus for $0<e<1$ we have that $\widehat{W}_r(e)>1$. 
This corresponds to the case when $b>a$. A similar analysis can be done for $a>b$ with a similar conclusion:
 the minimum  is at  $a=b$. In the other limit the  function $\widehat{W}_r$ approaches infinity for $b=0$ that corresponds to the case when one dimension of the ellipsoid 
shrinks to zero and it becomes a lower dimensional sphere.
This analysis demonstrates that in the class of ellipsoid geometries the round sphere is indeed the minimizer of the
normalized Willmore energy and, respectively, is the maximizer of the holographic entanglement entropy,
\be
S(E^{d-2})\leq S(S^{d-2})\, .
\lb{5.14}
\ee
This statement may be made more rigorous. Consider, a generic ellipsoid (\ref{5.6}) characterized by parameters $a_1, .., a_{d-1}$.
Both the area and the Willmore energy are symmetric functions of these parameters,
\be
A(E^{d-2})=A(a_1,..,a_{d-1})\, , \ \ W(E^{d-2})=W(a_1,.., a_{d-1})\, ,
\lb{5.15}
\ee
where $A(..)$ and $W(..)$ are symmetric functions of their arguments. Suppose that $a_{d-1}\neq 0$. Then, the normalized Willmore energy is a symmetric  function of 
$(d-2)$ variables
\be
\widehat{W}(E^{d-2})=\widehat{W}(\alpha_1,..,\alpha_{d-2})\, , \ \ \alpha_i=a_i/a_{d-1}\, .
\lb{5.16}
\ee
Suppose that this function has its minimum at values $\alpha_1^0$, .. ,$\alpha_{d-2}^0$. Then near this point in quadratic order  it can be 
presented as
\be
\widehat{W}(\alpha_1,..,\alpha_{d-2})=\sum_{i,j}W_{ij}(\alpha_i-\alpha_i^0)(\alpha_j-\alpha_j^0)\, ,
\lb{5.17}
\ee
where the condition of symmetry requires that $W_{ij}$ to be symmetric. This however is not sufficient for complete symmetry of (\ref{5.17}).
Indeed, interchanging any pair $\alpha_i$ and $\alpha_j$ we find that (\ref{5.17}) is symmetric only if $\alpha^0_j=\alpha_i^0$.
Thus, we conclude that the symmetry condition requires that $\alpha_1^0=\alpha_2^0=..=\alpha^0_{d-2}$. The respective ellipsoid geometry 
is precisely the round sphere. Since the minimum of the normalized Willmore energy corresponds to maximum of entanglement entropy we come to
inequality (\ref{5.14}).

\subsection{Product spaces $S^m\times S^n$}
In higher dimensions there are many possibilities to create a ``toric'' geometry by considering various products 
of spheres, $S^m\times S^n$.

A natural generalization for the flat metric of $R^{m+n+2}$, will be
\begin{equation}
\begin{split}
ds^2&=\bigg[dt^2+dr^2+r^2(d\theta_1^2+\sin^2\theta_1d\theta_2^2+\cdots +\sin^2\theta_1\cdots\sin^2\theta_{m-1}d\theta_m^2)\\
&+ (R+r\cos\theta_1)^2(d\alpha_1^2+\sin^2\alpha_1d\alpha_2^2+\cdots +\sin^2\alpha_1\cdots\sin^2\alpha_{n-1}d\alpha_n^2)
\bigg]\, ,
\end{split}
\lb{metric}
\end{equation}
Here, surface of constant $r$ is a product space as $S^m\times S^n$.
We can find the trace of the extrinsic curvature for such a general case, which reads
\begin{equation}
\Tr K(S^m\times S^n)=\frac{mR+(m+n)r\cos\theta_1}{r(R+r\cos\theta_1)}\, .
\end{equation}

The area of the surface $S^m\times S^n$ is found to be
\begin{equation}\label{area}
A\left(S^m\times S^n\right)=\int\prod_{i=1}^md\theta_i\prod_{j=1}^n d\alpha_j\,\, a(r,\theta_i,\alpha_j)\, ,
\end{equation}
where
\begin{equation}
\begin{split}
 a(r,\theta_i,\alpha_j)&=r^m\left[(\sin\theta_1)^{m-1}(\sin\theta_2)^{m-2}\cdots\sin\theta_{m-1}\right](R+r\cos\theta_1)^n \\
 &\times\left[(\sin\alpha_1)^{n-1}(\sin\alpha_2)^{n-2}\cdots\sin\alpha_{n-1}\right]\, .
\end{split}
\end{equation}
One can perform the integral \eqref{area} and find
\begin{equation}
A(S^m\times S^n)=\frac{4\pi^{\frac{m+n}{2}+1}}{\Gamma\left(\frac{m+1}{2}\right)\Gamma\left(\frac{n+1}{2}\right)}(R+r)^nr^m\,\, {_2F_1(\frac{m}{2},-n,m,\frac{2r}{R+r})}\, .
\end{equation}
We can also evaluate the  Willmore energy
\begin{equation}
W\left(S^m\times S^n\right)=\frac{1}{4}\int\prod_{i=1}^md\theta_i\prod_{j=1}^n d\alpha_j\,\,\left[\Tr K(S^m\times S^n)\right]^2\,a(r,\theta_i,\alpha_j)\, ,
\end{equation}
which leads to
\begin{equation}
\begin{split}
W\left(S^m\times S^n\right)&=\frac{\pi^{\frac{m+n}{2}+1}}{\Gamma\left(\frac{m+1}{2}\right)\Gamma\left(\frac{n+1}{2}\right)}(R+r)^nr^{m-2}\bigg[(m+n)^2\,\,{_2F_1(\frac{m}{2},-n,m,\frac{2r}{R+r})}\\
&+\frac{n^2R^2}{(R+r)^2}\,\,{_2F_1(\frac{m}{2},-n+2,m,\frac{2r}{R+r})}\\
&-\frac{2(m+n)nR}{R+r}\,\,{_2F_1(\frac{m}{2},-n+1,m,\frac{2r}{R+r})}\bigg]\, .
\end{split}
\end{equation}
Some specific examples read
\begin{equation}
\begin{split}
&m=1,n=1\,\,\rightarrow\,\, A\left(S^1\times S^1\right)=4\pi^2rR\,\,\,\, , \,\,\,\, W\left(S^1\times S^1\right)=\frac{\pi^2R^2}{r\sqrt{R^2-r^2}}\, ,\\
&m=2,n=1\,\,\rightarrow\,\, A\left(S^2\times S^1\right)=8\pi^2r^2R\,\,\,\, , \,\,\,\, W\left(S^2\times S^1\right)=\frac{\pi^2R}{r}\left[6r+R\log\left(\frac{R+r}{R-r}\right)\right]\, ,\\
&m=1,n=2\,\,\rightarrow\,\, A\left(S^1\times S^2\right)=4\pi^2r(r^2+2R^2)\,\,\,\, , \,\,\,\, W\left(S^1\times S^2\right)=\frac{\pi^2}{r}(9r^2+2R^2)\, ,
\end{split}
\end{equation}
and so on.

Now defining $x=\frac{r}{R}$ we can construct the following dimensionless quantity
\begin{equation}\label{Wb1}
\begin{split}
\widehat{W}\left(S^m\times S^n\right)&=\frac{W\left(S^m\times S^n\right)}{\left[A\left(S^m\times S^n\right)\right]^\frac{m+n-2}{m+n}}=\\
&\frac{1}{4^{\frac{m+n-2}{m+n}}}\left(\frac{\pi^{\frac{m+n}{2}+1}}{\Gamma\left(\frac{m+1}{2}\right)\Gamma\left(\frac{n+1}{2}\right)}\right)^{\frac{2}{m+n}}(1+\frac{1}{x})^\frac{2n}{m+n}\times\bigg[(m+n)^2\,\,{_2F_1(\frac{m}{2},-n,m,\frac{2x}{1+x})}\\
&+\frac{n^2}{(1+x)^2}\,\,{_2F_1(\frac{m}{2},-n+2,m,\frac{2x}{1+x})}-\frac{2(m+n)n}{1+x}\,\,{_2F_1(\frac{m}{2},-n+1,m,\frac{2x}{1+x})}\bigg]\\
&\times\bigg[{_2F_1(\frac{m}{2},-n,m,\frac{2x}{1+x})}\bigg]^{-\frac{m+n-2}{m+n}}\, .
\end{split}
\end{equation}

Our desired problem is to compare the normalized Willmore energy  of surface  $S^m\times S^n$  with that of a round sphere with the same fixed area.
Therefore, we shall consider a ratio of two normalized energies
\be
\widehat{W}_r(x)=\frac{\widehat{W}(S^m\times S^n)}{\widehat{W}(S^{m+n})}\, .
\lb{WW}
\ee
as function of the variable $x=r/R$ and will look for a minimum of this function. Notice that $0\leq x\leq 1$ as one can see from the metric (\ref{metric}).
In what follows we have calculated the minimum value for this quantity in $d=4,5,\dots,10$ dimensions.

\begin{itemize}
\item $d=m+n+2=4$
\begin{center}
\begin{tabular}{| c | c | }
  \hline                       
  d=4 & $S^1\times S^1$  \\
  \hline
  $x_{min}$ & 0.707  \\
 $\widehat{W}_{r,min}$ & 1.571 \\
  \hline  
\end{tabular}
\end{center}

\item $d=m+n+2=5$
\begin{center}
\begin{tabular}{| c | c | c | }
  \hline                       
  d=5 & $S^2\times S^1$ & $S^1\times S^2$ \\
  \hline
  $x_{min}$ & 0.886 & 0.816  \\
 $\widehat{W}_{r,min}$ & 1.391 & 1.333 \\
  \hline  
\end{tabular}
\end{center}

\item $d=m+n+2=6$
\begin{center}
\begin{tabular}{| c | c | c | c | }
  \hline                       
  d=6 & $S^3\times S^1$ & $S^2\times S^2$ & $S^1\times S^3$ \\
  \hline
  $x_{min}$ & 0.968 & 1 & 1 \\
 $\widehat{W}_{r,min}$ & 1.324 & 1.237 & 1.116\\
  \hline  
\end{tabular}
\end{center}

\item $d=m+n+2=7$
\begin{center}
\begin{tabular}{| c | c | c | c | c | }
  \hline                       
  d=7 & $S^4\times S^1$ & $S^3\times S^2$ & $S^2\times S^3$ & $S^1\times S^4$ \\
  \hline
  $x_{min}$ & 0.9987 & 1 & 1 & 1 \\
 $\widehat{W}_{r,min}$ & 1.289 & 1.226 & 1.152 & 1.076\\
  \hline  
\end{tabular}
\end{center}

\item $d=m+n+2=8$
\begin{center}
\begin{tabular}{| c | c | c | c | c | c |}
  \hline                       
  d=8 & $S^5\times S^1$ & $S^4\times S^2$ & $S^3\times S^3$ & $S^2\times S^4$ & $S^1\times S^5$ \\
  \hline
  $x_{min}$ & 1 & 1 & 1 & 1 & 1 \\
 $\widehat{W}_{r,min}$ & 1.271 & 1.230 & 1.175 & 1.117 & 1.058\\
  \hline  
\end{tabular}
\end{center}

\item $d=m+n+2=9$
\begin{center}
\begin{tabular}{| c | c | c | c | c | c | c |}
  \hline                       
  d=9 & $S^6\times S^1$ & $S^5\times S^2$ & $S^4\times S^3$ & $S^3\times S^4$ & $S^2\times S^5$  & $S^1\times S^6$\\
  \hline
  $x_{min}$ & 1 & 1 & 1 & 1 & 1 & 1 \\
 $\widehat{W}_{r,min}$ & 1.257 & 1.233 & 1.192 & 1.145 & 1.097 & 1.048\\
  \hline  
\end{tabular}
\end{center}

\item $d=m+n+2=10$
\begin{center}
\begin{tabular}{| c | c | c | c | c | c | c | c |}
  \hline                       
  d=10 & $S^7\times S^1$ & $S^6\times S^2$ & $S^5\times S^3$ & $S^4\times S^4$ & $S^3\times S^5$  & $S^2\times S^6$ & $S^1\times S^7$\\
  \hline
  $x_{min}$ & 1 & 1 & 1 & 1 & 1 & 1 & 1 \\
 $\widehat{W}_{r,min}$ & 1.245 & 1.234 & 1.204 & 1.165 & 1.124 & 1.083 & 1.041\\
  \hline  
\end{tabular}
\end{center}
\end{itemize}

\bigskip

\begin{figure}
\centering
  \includegraphics[width=0.32\textwidth]{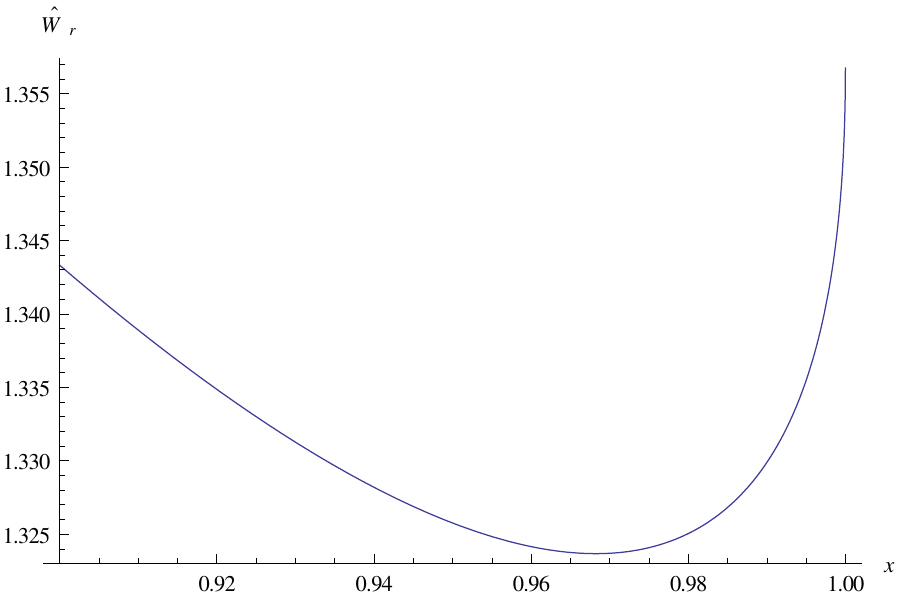}
  \includegraphics[width=0.32\textwidth]{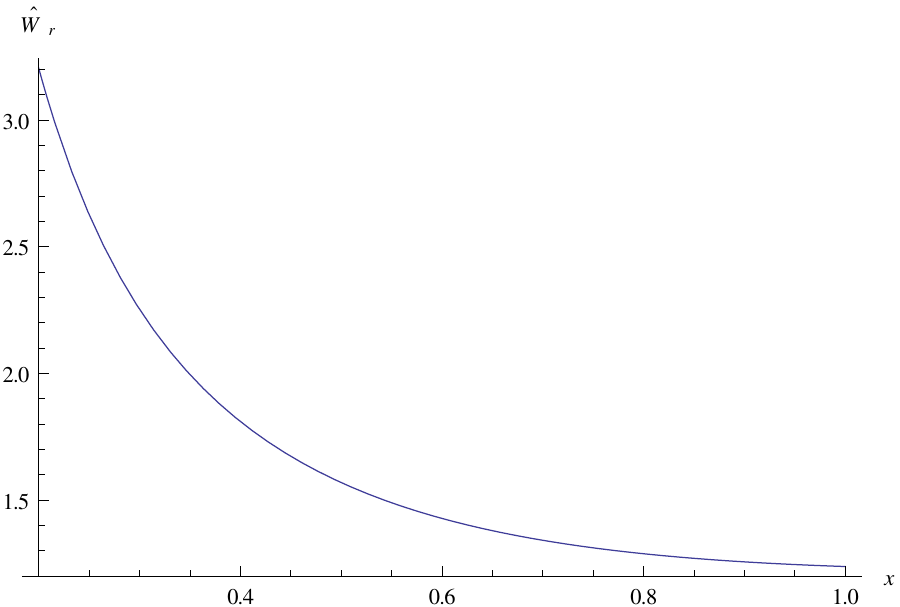}
    \includegraphics[width=0.32\textwidth]{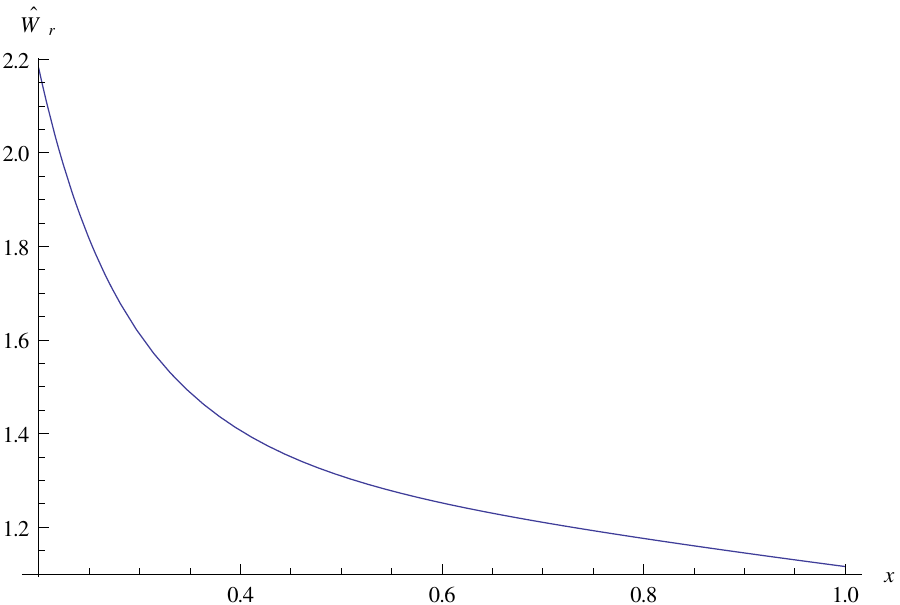}
  \caption{$\widehat{W}_r(x)$ for (from left to right) $S^3\times S^1$, $S^2\times S^2$ and $S^1\times S^3$, respectively.}\label{fig2}
\end{figure}

Based on these data one can make the following observations:
\begin{itemize}
\item[1)]
Only in  five cases, i.e. $S^1\times S^1$, $S^1\times S^2$, $S^2\times S^1$, $S^3\times S^1$ and $S^4\times S^1$, is the local minimum 
  at $x_{min}<1$. For other cases the function 
$\widehat{W}_{r}(x)$ is monotonically decreasing so that the minimum  occurs at $x_{min}=1$. For illustrative purposes we present the different 
behavior of the  function 
$\widehat{W}_{r}(x)$ in Figure (\ref{fig2}).
\item[2)]
For all combinations of $m$ and $n$, we have $\widehat{W}_{r,min}>1$. This means that the Willmore energy for $S^m\times S^n$ topology is greater than  the Willmore energy for the round sphere of same dimension, i.e. $S^{m+n}$. So that in comparison with a general toric geometry, the round spheres are the maximizers of the entanglement entropy.
\item[3)]
In each table, $\widehat{W}_{r,min}$ decreases from left to right. Consequently, for spaces $S^m\times S^n$  in $d=m+n+2$ dimension, entanglement entropy will have its maximum value for the space $S^1\times S^{d-3}$. 
\item[4)] It also decreases from the top to the bottom. In particular, in the limit  $d\rightarrow\infty$ we have  $\widehat{W}_{r,min}\rightarrow 1$.
In this limit entanglement the entropy thus approaches that of the round sphere.
\end{itemize}
\medskip

We can also formulate some conclusions/conjectures:
\begin{itemize}
\item[1)] In the class of compact $(d-2)$-surfaces of arbitrary topology the round sphere $S^{m+n}$, $m+n=d-2$, is the global minimizer of the
normalized Willmore energy. Respectively, it is the global maximizer of the entanglement entropy in agreement with our guess in section 2.3.
\item[2)] For surfaces of fixed topology same as that of $S^m\times S^n$ the product space $S^m\times S^n$ (with the radii ratio $x_{min}$ given in the tables above) is the  minimizer 
\be
\widehat{W}(\Sigma)\geq \widehat{W}_{\rm min}(S^m\times S^n)\, , 
\ee
where the minimal values are given in the tables above. Respectively, for surfaces of  this topology the entanglement entropy
satisfies the bound
\be
S(\Sigma)\leq S(S^m\times S^n)\, .
\lb{SS}
\ee
\end{itemize}

\section{Conclusions}

In this paper we have analyzed the problem of finding a surface for which the entanglement entropy of a given quantum field theory would be maximal.
In four dimensions for a large class of conformal and non-conformal field theories this problem reduces to a well-known mathematical problem
of minimization of the Willmore bending energy. In each topological class there exists a surface, known as the Lawson surface, which minimizes the 
Willmore energy and respectively maximizes the entanglement entropy. The global maximizer (in all possible topologies) is the round sphere.
This fact may have some important applications in  various physical 
 models and situations and may be used as a hint to actually observe the entanglement entropy
in an experiment. We however do not dwell into this problem here. 

In higher dimensions the analysis is somewhat more involved. In particular, there are more possibilities to  construct surfaces of ``toric'' type, namely
the product spaces $S^m\times S^n$ with various possible values for $n$ and $m$ such that $n+m+2=d$ is dimension of the physical spacetime.
We have conjectured that these product spaces are the entropy maximizers in their own topological class. On the other hand, our analysis indicates that the global maximizer
in all possible topologies is still the round sphere. This fact appears to be universal in all  dimensions $d\geq 4$. 

The dimension $d=3$ needs a special consideration. 
The reason is that in this dimension the first subleading term in 
the entropy is a constant which does not depend on the UV cut-off.
Therefore in order to calculate this constant and determine its sign, 
in the holographic set up, we would need to know the 
respective minimal surface and its area
exactly, not making any approximations.  
This may be a difficult task if the entangling surface 
(which has  dimension 1 in this case) is some arbitrary closed curve.
Thus we have not done this analysis in dimension $d=3$. 
However, we expect that the round circle $S^1$ 
would still be the global maximizer in this case.
It would be nice to check this conjecture in some explicit calculation.

We are not aware of any previous mathematical results on 
the Willmore energy in dimensions higher than 4. 
Therefore, we expect that our conclusions and conjectures
made in section 4 may have some mathematical significance.

\section*{Acknowledgements} 
SS would like to thank the Theory Division at CERN and the Yukawa Institute for Theoretical Physics (Kyoto) for kind hospitality during this project.
SS thanks Tadashi Takayanagi for useful discussions. 
 AF would like to thank  Laboratoire de Math\'ematiques et Physique Th\'eorique (LMPT) for hospitality during this work. AF is also grateful to Hessamaddin Arfaei and Amir Esmaeil Mosaffa for encouragement and support.
G.W.G. would like to record his gratitude for the kindness and hospitality
shown to him  by the late Tom Wilmore when, as a research student,
he consulted him on matters related to the 
Penrose inequality  and  for informing
him at that time   of the Wilmore  conjecture.

\end{document}